\documentclass[a4paper,fleqn,usenatbib,useAMS]{mnras}

\usepackage{mathptmx}

\usepackage[T1]{fontenc}
\usepackage{ae,aecompl}

\usepackage{graphicx,epstopdf}	
\usepackage{amsmath}	
\usepackage{amssymb}	
\usepackage{mathtools}
\usepackage{upgreek}

\defcitealias{Ginzburg2016}{GSS16}

\title[Core-powered mass loss]{Core-powered mass loss and the radius distribution of small exoplanets}

\author[S. Ginzburg, H. E. Schlichting and R. Sari]{
Sivan Ginzburg,$^{1}$\thanks{E-mail: sivan.ginzburg@mail.huji.ac.il}
Hilke E. Schlichting$^{2,3}$
and Re'em Sari$^{1}$
\\
$^{1}$Racah Institute of Physics, The Hebrew University, Jerusalem 91904, Israel\\
$^{2}$Department of Earth, Planetary, and Space Sciences, University of California, Los Angeles, CA 90095, USA\\
$^{3}$Department of Earth, Atmospheric and Planetary Sciences, Massachusetts Institute of Technology, Cambridge, MA 02139, USA}

\date{Accepted XXX. Received YYY; in original form ZZZ}

\pubyear{2017}

\begin{document}
\label{firstpage}
\pagerange{\pageref{firstpage}--\pageref{lastpage}}
\maketitle

\begin{abstract}
Recent observations identify a valley in the radius distribution of small exoplanets, with planets in the range $1.5-2.0\,{\rm R}_{\earth}$ significantly less common than somewhat smaller or larger planets. This valley may suggest a bimodal population of rocky planets that are either engulfed by massive gas envelopes that significantly enlarge their radius, or do not have detectable atmospheres at all. One explanation of such a bimodal distribution is atmospheric erosion by high-energy stellar photons. We investigate an alternative mechanism: the luminosity of the cooling rocky core, which can completely erode light envelopes while preserving heavy ones, produces a deficit of intermediate sized planets. We evolve planetary populations that are derived from observations using a simple analytical prescription, accounting self-consistently for envelope accretion, cooling and mass loss, and demonstrate that core-powered mass loss naturally reproduces the observed radius distribution, regardless of the high-energy incident flux. Observations of planets around different stellar types may distinguish between photoevaporation, which is powered by the high-energy tail of the stellar radiation, and core-powered mass loss, which depends on the bolometric flux through the planet's equilibrium temperature that sets both its cooling and mass-loss rates.  
\end{abstract}




\section{Introduction}\label{sec:introduction}

Over the past decade, the {\it Kepler} mission has revealed an ubiquitous population of short-period planets smaller than Neptune \citep[$\lesssim 4\,{\rm R}_{\earth}$; see, e.g.,][and references therein]{Marcy2014}. Combining the transit radius with mass estimates from either radial velocity measurements or transit timing variations provides a handle on the density of some of these planets, indicating that many of them do not have a rocky composition \citep{WeissMarcy2014,JontofHutter2016,HaddenLithwick2017}. A commonly suggested structure that explains the low densities of these planets is a rocky core engulfed by a voluminous gas atmosphere of several percent in mass \citep[e.g.][]{Lopez2012,Lissauer2013,WolfgangLopez2015}. Theoretical models demonstrate that such atmospheres can be gravitationally accreted by the cores from the gaseous nebulae that surround young stars for their first few million years \citep{Lee2014,InamdarSchlichting2015,LeeChiang2015,Ginzburg2016}. 

Recently, refined radius measurements by the California-Kepler Survey \citep{Johnson2017,Petigura2017} allowed \citet{Fulton2017} to identify a distinct valley in the radius distribution of short-period planets. Explicitly, planets with a radius of $\approx 1.8\,{\rm R}_{\earth}$ are significantly less common than smaller ($\approx 1.3\,{\rm R}_{\earth}$) or larger ($\approx 2.4\,{\rm R}_{\earth}$) planets. A plausible interpretation of this valley is that planets either come as bare rocky cores (the low-radius peak) or with gas envelopes that are massive enough to roughly double their radius (the high-radius peak).

\cite{OwenWu13} have already noticed such a valley in the observations, and explained it by atmosphere evaporation due to high-energy stellar photons \citep[see also][]{LopezFortney13,Jin2014,ChenRogers2016,LehmerCatling2017}. In a recent paper, \citet{OwenWu2017} use analytical arguments to illustrate how photoevaporation naturally sculpts a bimodal distribution that resembles the results of \citet{Fulton2017}.  

Here, we focus on an alternative mechanism that is unrelated to the high-energy flux and yet also naturally reproduces the observed bimodal distribution. In a previous paper \citep[][hereafter \citetalias{Ginzburg2016}]{Ginzburg2016} we showed that, following the dispersal of the protoplanetary disc, the heat from the cooling inner layers of the atmosphere unbinds the loosely bound outer layers, on a timescale of a few million years. As a result, the planet's radius shrinks to roughly twice the size of the rocky core in this time \citep[see also][]{OwenWu2016}. The fate of the gas envelope from this stage onward is determined by the ratio of its heat capacity to that of the rocky core. In \citetalias{Ginzburg2016} we explained that if the core dominates the heat capacity then its cooling luminosity might strip off the overlying atmosphere entirely, leaving behind a bare core. More massive atmospheres, on the other hand, survive and cool without being affected by the core's luminosity. Since massive envelopes are unaffected, while light ones can be lost completely, core-powered mass loss naturally leads to a bimodal distribution of final atmosphere masses and planet radii.   

In this paper we implement the theory presented in \citetalias{Ginzburg2016} and demonstrate that core cooling can reproduce the observational valley in \citet{Fulton2017}. 

\section{Planet Structure and Evolution}\label{sec:evolution}

In this section we briefly review the core plus envelope structure and evolution of short-period planets a few times the mass of Earth (${\rm M}_{\earth}$) after the dispersal of the protoplanetary disc, with a detailed description given in \citetalias{Ginzburg2016}.

We assume a rocky core\footnote{Arguments similar to those in \citet{JinMordasini2017} can be used to distinguish between rocky and icy cores scenarios.} of mass $M_c$ and radius $R_c$, that are related by $M_c/{\rm M}_{\earth}=(R_c/{\rm R}_{\earth})^4$ \citep[due to gravitational compression; see][]{Valencia2006}. 
The core is engulfed by a gaseous atmosphere, with a mass fraction $f\equiv M_{\rm atm}/M_c$, that can be divided into a radiative (nearly isothermal) envelope with negligible mass and a convective interior, connected at the radiative--convective boundary, $R_{\rm rcb}$.
In \citetalias{Ginzburg2016} we showed that the dispersal of the protoplanetary disc causes a loss of pressure support that triggers atmospheric mass loss (powered by the luminosity of the cooling inner layers) until the atmosphere shrinks to a thickness $\Delta R\equiv R_{\rm rcb}-R_c\approx R_c$ (note the slight notation change compared to \citetalias{Ginzburg2016}: here, $\Delta R$ marks the thickness of the convective layer, whereas $R_{\rm rcb}$ is the radius of the RCB, measured from the planet's centre) in a few million years ($R_{\rm rcb}$ is a good approximation for the planet's observed radius, since the scale height of the radiative layer is much smaller than $R_c$; see \citetalias{Ginzburg2016}).

Here, we focus on the thin-atmosphere regime ($\Delta R\lesssim R_c$) since the core does not cool during the preceding thick regime, as we explain below. More precisely, as explained in \citetalias{Ginzburg2016}, atmospheres transition smoothly from the thick ($1\lesssim\Delta R/R_c\lesssim R_{\rm B}/R_c$) to the thin ($R_c/R_{\rm B}<\Delta R/R_c\lesssim 1$) and finally to the ultra thin ($\Delta R/R_c\lesssim R_c/R_{\rm B}$) regime, with $R_{\rm B}\sim GM_c\mu/(k_{\rm B}T_{\rm eq})$ denoting the Bondi radius ($G$, $\mu$, $k_{\rm B}$ and $T_{\rm eq}$ are defined below). As described in Section \ref{sec:obs}, the observed valley in the radius distribution is at $R_c/R_{\rm B}\lesssim \Delta R/R_c\lesssim 1$, further encouraging us to focus on the thin regime.

By solving the hydrostatic equilibrium for an ideal gas in the convective layer (where the pressure scales with density as $P\propto\rho^{\gamma}$) we can calculate the temperature, density, and pressure profiles of the atmosphere. Specifically, by integrating the density profile, we find that the mass of the atmosphere is given by 
\begin{equation}\label{eq:matm_thin}
	M_{\rm atm}=\frac{\gamma-1}{\gamma}4\uppi R_c^2\rho_{\rm rcb}\Delta R\left(\frac{R_{\rm B}'\Delta R}{R_c^2}\right)^{1/(\gamma-1)}.
\end{equation}
Here, $\gamma$ denotes the adiabatic index, $\rho_{\rm rcb}$ is the density at the RCB, and $R_{\rm B}'$ is the modified Bondi radius:
\begin{equation}\label{eq:bondi}
	R_{\rm B}'\equiv\frac{\gamma-1}{\gamma}\frac{GM_c\mu}{k_{\rm B}T_{\rm rcb}},
\end{equation}
where $G$ is the gravitational constant, $\mu$ is the atmosphere's molecular mass, $k_{\rm B}$ is Boltzmann's constant, and $T_{\rm rcb}\sim T_{\rm eq}$ is the temperature at the RCB, which is similar, for the power-law opacities that we incorporate (Section \ref{sec:impl}), to the equilibrium temperature (determined by the distance from the star). 

Similarly, by calculating the temperature profile, we find the temperature at the bottom of the atmosphere ($R=R_c$)
\begin{equation}\label{eq:t_c}
k_{\rm B}T_c=\frac{\gamma-1}{\gamma}\frac{GM_c\mu}{R_c^2}\Delta R,
\end{equation} 
which is valid for $R_c/R_{\rm B}'\lesssim \Delta R/R_c\lesssim 1$. The ultra thin regime ($\Delta R/R_c<R_c/R_{\rm B}'\approx 0.1$, for which $T_c\sim T_{\rm eq}$) is relevant only for $\Delta R\lesssim 0.1 R_c$ and therefore does not affect the observed radius distribution. We treat the rocky core as a roughly isothermal ball with a temperature $T_c$, given by equation \eqref{eq:t_c}. This approximation relies on the assumption that the core is roughly incompressible, molten, and therefore fully convective. While we mention in \citetalias{Ginzburg2016} the possibility of forming an insulating solid crust, it is easy to verify, using equation \eqref{eq:t_c}, that our typical cores ($M_c\approx 3M_{\earth}$; see Section \ref{sec:results}) cool down to the rock melting temperature only when their atmospheres are as thin as $\Delta R\lesssim 0.1 R_c$. Therefore, crust formation does not significantly affect the shape of the planet radius distribution. We note, however, that the core-convection might occur on a slow timescale if the molten rock's viscosity is high enough \citep{Stamenkoic2012}. We disregard this possibility here for simplicity and assume that the bottleneck for the convective cooling of both the core and the atmosphere is the diffusion through the atmosphere's radiative layer. Explicitly, the planet cools with a luminosity
\begin{equation}\label{eq:lum}
	L=-\dot{E}_{\rm cool}=\frac{64\uppi}{3}\frac{\sigma T_{\rm rcb}^4R_{\rm B}'}{\kappa\rho_{\rm rcb}},
\end{equation}
where $\sigma$ is the Stefan-Boltzmann constant and $\kappa$ is the opacity at the RCB. 

The available energy for cooling (gravitational and thermal) is given by
\begin{equation}\label{eq:enr_thin}
	E_{\rm cool}=g\Delta R\left(\frac{\gamma}{2\gamma-1}M_{\rm atm}+\frac{1}{\gamma}\frac{\gamma-1}{\gamma_c-1}\frac{\mu}{\mu_c}M_c\right),
\end{equation} 
where $\mu_c$ and $\gamma_c$ mark the rocky core's molecular weight and adiabatic index, respectively, and $g\equiv GM_c/R_c^2$ is the surface gravity. The first term in equation \eqref{eq:enr_thin} is the energy of the gaseous envelope (obtained by integrating its density and temperature profiles), while the second term is the thermal energy of the rocky core. This term is obtained by assuming that the core is at a uniform temperature that is given by equation \eqref{eq:t_c}, as discussed above. Equation \eqref{eq:enr_thin} emphasizes a fundamental difference between thin ($\Delta R\lesssim R_c$) and thick ($\Delta R\gtrsim R_c$) envelopes. The temperature at the bottom of a thick convective envelope is an approximately constant $k_{\rm B}T_c=(\gamma-1)/\gamma\times GM_c\mu/R_c$ (see \citetalias{Ginzburg2016}), regardless of $\Delta R$. Therefore, as thick envelopes cool and contract, their underlying rocky cores remain almost at the same temperature and do not contribute significantly to the heat capacity. Only once the envelopes reach the thin regime (which is the focus of this work), the cores beneath them can cool, giving rise to the second term in the equation.

Quantitatively, the ratio between the heat capacities of the core and the envelope is of the order of $(\mu/\mu_c)f^{-1}$ in the thin regime, as evident from equation \eqref{eq:enr_thin}. In the preceding thick regime ($\Delta R>R_c$), on the other hand, this ratio is given by a smaller $(\mu/\mu_c)(R_c/\Delta R)^{1/2}f^{-1}$ for the same atmosphere mass fraction $f$, as can be derived from equations (5) and (12) in \citetalias{Ginzburg2016} (for $\gamma=7/5$ which we choose below).

\subsection{Spontaneous mass loss}\label{sec:evaporation}

As a result of the vanishing pressure boundary condition at infinity once the nebula disperses, gas tends to escape if there is enough energy to lift it out from the planet's potential well \citepalias[see][]{Ginzburg2016}. In our case, the cooling luminosity $L=-\dot{E}_{\rm cool}$, given by equation \eqref{eq:lum}, may provide this energy. We thus identify \citep[see also][]{IkomaHori2012,OwenWu2016} a process of spontaneous mass loss that is essentially different from (and complementary to) the high-energy photoevaporation that is usually discussed in the literature \citep{MurrayClay2009,Lopez2012,OwenJackson2012, OwenWu13}. The energy required to blow off the entire atmosphere is $E_{\rm loss}=GM_cM_{\rm atm}/R_c=M_{\rm atm}g R_c$. By comparing $E_{\rm loss}$ to $E_{\rm cool}$ from equation \eqref{eq:enr_thin} we learn that once atmospheres reach the thin regime, their fate is determined by the ratio of their heat capacity to that of the core. 

\subsubsection{Heavy Atmospheres}

If $M_{\rm atm}/M_c>\mu/\mu_c\sim 5\%$ (we temporarily omit the $\gamma$ and $\gamma_c$ factors for simplicity) then initially (when $\Delta R=R_c$) $E_{\rm cool}\sim E_{\rm loss}$ and the atmosphere cools and loses mass at the same rate. However, once the atmosphere shrinks (due to cooling), $\Delta R<R_c$ and $E_{\rm cool}<E_{\rm loss}$. Therefore, the cooling time of heavy envelopes is shorter than their mass-loss time and they survive roughly intact (cooling without losing mass, with an analytical solution given in \citetalias{Ginzburg2016}).

\subsubsection{Light Atmospheres}\label{sec:light}  

If $M_{\rm atm}/M_c<\mu/\mu_c\sim 5\%$ then the heat capacity is dominated by the rocky core and $E_{\rm cool}>E_{\rm loss}$, implying that the mass-loss timescale is shorter than the cooling time. As the mass of the atmosphere decreases, the mass-loss time becomes even shorter, resulting in a rapid removal of the entire envelope. Explicitly, since $E_{\rm cool}\propto M_c\Delta R$ according to equation \eqref{eq:enr_thin}, the atmosphere loses mass primarily by decreasing its density (and $\rho_{\rm rcb}$), according to equation \eqref{eq:matm_thin}, whereas $\Delta R$ and $E_{\rm cool}$ change more slowly (maintaining $E_{\rm cool}>E_{\rm loss}$). In other words, the shrinking of the envelope is (to leading order) inhibited by the core's heat, and it loses mass faster than it shrinks. We note, however, that spontaneous mass loss is also limited by the escape rate of molecules at the speed of sound $c_s\simeq(k_{\rm B} T_{\rm eq}/\mu)^{1/2}$ at the Bondi radius $R_{\rm B}\equiv GM_c\mu/(k_{\rm B}T_{\rm eq})\gg R_c$ \citep[see also][]{OwenWu2016}\footnote{The Hill radius $R_{\rm H}$ may also set a boundary condition. However, for short-period super Earths, $R_{\rm H}\sim R_{\rm B}$ \citep{Lee2014,InamdarSchlichting2015}} 
\begin{equation}\label{eq:bondi_condition}
	|\dot{M}_{\rm atm}|<{\dot M}_{\rm atm}^{\rm B}\equiv 4\uppi R_{\rm B}^2\rho(R_{\rm B})c_s,
\end{equation}
where we relate the density at the Bondi radius to that at the RCB using hydrostatic equilibrium in the nearly isothermal radiative layer
\begin{equation}\label{eq:rho_rb}
	\frac{\rho(R_{\rm B})}{\rho_{\rm rcb}}=\exp\left(-\frac{R_{\rm B}}{R_{\rm rcb}}+1\right).
\end{equation}
In \citetalias{Ginzburg2016} we demonstrated that the exponential dependence on $R_{\rm B}/R_{\rm rcb}\propto M_c^{3/4}/T_{\rm eq}$ in equation \eqref{eq:rho_rb} differentiates between light or hot planets that lose their atmospheres within a few gigayears, and massive or cold ones that partially retain their gas envelopes. It is this additional limit on the mass-loss rate that allows some atmospheres with $M_{\rm atm}/M_c<5\%$ to avoid complete removal in a runaway process, producing some (relatively rare) planets with $0<M_{\rm atm}/M_c<5\%$. 

By definition, $R_{\rm rcb}$ does not change significantly during the thin regime, which is the focus of this work. Therefore, the limit on the mass-loss timescale posed by equation \eqref{eq:rho_rb} remains approximately constant as the planet cools (in the numerical implementation we take the exact $R_{\rm rcb}=R_c+\Delta R(t)$; see Section \ref{sec:impl}).
The situation is different during the earlier thick regime, when the atmosphere contracts significantly due to cooling. During that regime (which is not discussed in this work) the competition between the cooling time and the disc dispersal timescale determines which atmospheres suffer from significant mass loss \citep[see Section 2.2 of][]{OwenWu2016}.

\subsection{Implementation}\label{sec:impl}

We implement the simultaneous cooling and mass loss (spontaneous loss due to cooling; we disregard photoevaporation for simplicity and to separate these two different effects) by keeping track of the planet's energy $E_{\rm cool}$ and envelope mass $M_{\rm atm}$. In each time step $\Delta t$ we evolve the planet:
\begin{subequations}
	\begin{equation}\label{eq:cooling_eq}
		E_{\rm cool}\to E_{\rm cool}-L\Delta t
	\end{equation}
	\begin{equation}\label{eq:mass_loss_eq}
		M_{\rm atm}\to M_{\rm atm}-\min\left(\frac{L}{gR_c},{\dot M}_{\rm atm}^{\rm B}\right)\Delta t,
	\end{equation}	
\end{subequations}
with the luminosity $L$ and the Bondi rate ${\dot M}_{\rm atm}^{\rm B}$ given by equations \eqref{eq:lum} and \eqref{eq:bondi_condition}. More accurately, the mass-loss efficiency is not 100\%, as implied by equation \eqref{eq:mass_loss_eq}, since roughly half of the luminosity is radiated away when $L<{\dot M}_{\rm atm}^{\rm B}gR_c$ (required to sustain the radiative--convective profile), introducing an order of unity coefficient which we omit. 

The order of unity efficiency can also be understood by the following argument. As the pressure boundary condition is removed, gas escapes from the Bondi radius, initially at the Bondi rate, given by equation \eqref{eq:bondi_condition}. However, once the small amount of gas near the Bondi radius escapes, gas from deeper in the planet's potential well (near $R_{\rm rcb}$) must expand towards $R_{\rm B}$ in order for the mass loss to continue. Due to adiabatic cooling during this expansion, heat must be supplied in order for the gas to reach $R_{\rm B}$ \citepalias[see][]{Ginzburg2016}. In our case, this heat is supplied by the planet's cooling luminosity $L$, at the expanse of the radiated luminosity. However, as the radiated luminosity decreases, the RCB is pushed inward, lowering the Bondi mass-loss rate, according to equations \eqref{eq:bondi_condition} and \eqref{eq:rho_rb}. This reduction in the mass loss rate continues until the luminosity required to lift the escaping mass from the planet's potential well, $gR_c{\dot M}_{\rm atm}$, becomes significantly smaller ($\sim 1/2$) than the planet's cooling luminosity $L$, so that the radiated luminosity $L-gR_c{\dot M}_{\rm atm}$ does not change significantly.

In each time step, the time interval $\Delta t$ is chosen to be one percent of the minimum between the current cooling and mass-loss timescales. At the end of each time step, the envelope's density ($\rho_{\rm rcb}$) and thickness ($\Delta R$) are updated according to equations \eqref{eq:matm_thin} and \eqref{eq:enr_thin}. Using this procedure we obtain the planet's radius ($R_c+\Delta R$) and the atmosphere's mass as a function of time.

Following \citet{Freedman2008}, the (combination of molecular and alkali) opacity at the RCB is given by (valid for our temperature range $500\textrm{ K}<T_{\rm rcb}<2000\textrm{ K}$)
\begin{equation}
	\frac{\kappa}{0.1\textrm{ cm}^2\textrm{ g}^{-1}}=\left(\frac{\rho_{\rm rcb}}{10^{-3}\textrm{ g cm}^{-3}}\right)^{0.6}.
\end{equation}
The adiabatic index is the diatomic\footnote{See, however, the discussion in \citet{Lee2014}, \citet{LeeChiang2015}, and \citet{Piso2015}. Different choices of $\gamma$ can significantly alter the structure and evolution of thick envelopes ($\Delta R\gtrsim R_c$), but not of thin envelopes, on which we focus here.}
$\gamma=7/5$ (the molecular mass $\mu$ is also that of diatomic Hydrogen), and for the sake of simplicity and comparison with \citet{OwenWu2017} we assume that the ratio of the envelope's to the core's heat capacities is simply $17f$, with a similar result obtained by substituting $\gamma_c$ and $\mu_c$ in equation \eqref{eq:enr_thin}. Consequently, envelopes with mass fractions $f\gtrsim 5\%$ regulate their own cooling and survive, while lighter envelopes are dominated by the heat from the underlying rocky core and may be blown off entirely (see Section \ref{sec:evaporation}).

\section{Results}\label{sec:results}

In Section \ref{sec:evolution} we intuitively explained how core cooling might naturally lead to a bimodal planet population, as described by \citet{Fulton2017}. Here, we evolve planets according to the implementation given in Section \ref{sec:impl} and compare the resulting distributions to the observations.

\subsection{Simple demonstration}\label{sec:simple}

\begin{figure}
	\includegraphics[width=\columnwidth]{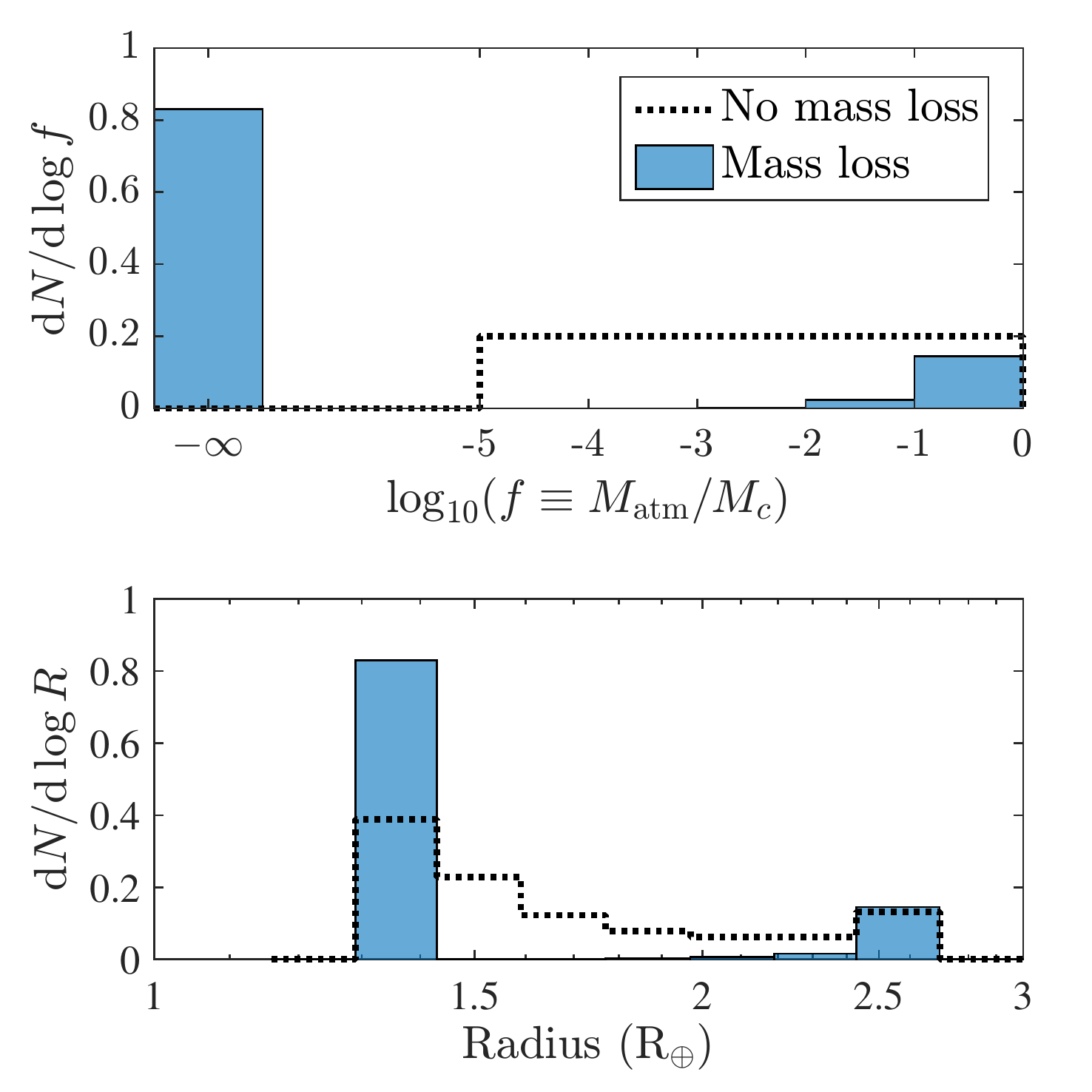}
	\caption{Distribution of the atmosphere mass fraction $f$ (top panel) and of the planet radius $R$ (bottom panel) after 3 gigayears of evolution with (blue bars) and without (dotted black line) mass loss (i.e. cooling only). All planets have a core mass $M_c=3{\rm M}_{\earth}$, equilibrium temperature $T_{\rm eq}\sim T_{\rm rcb}=10^3\textrm{ K}$ and start from twice the radius of the core $R=2R_c$. The initial atmosphere mass fractions are distributed logarithmically flat (top panel, dotted black line). The left blue column represents planets that have lost their atmospheres entirely due to core cooling.}
	\label{fig:simple}
\end{figure}

In Fig. \ref{fig:simple} we demonstrate how core-powered mass loss transforms an initially uniform distribution of $f$ into a bimodal population of atmosphere masses. We evolved 1000 planets for 3 gigayears, all with $M_c=3{\rm M}_{\earth}$ (motivated by the observed peak at $R\approx 1.3{\rm R}_{\earth}$), $T_{\rm rcb}=10^3\textrm{ K}$, and starting from $\Delta R=R_c$ (all planets reach this atmosphere thickness after a few million years, see \citetalias{Ginzburg2016}, so all the planets in Fig. \ref{fig:simple} start from $R=2.6\,{\rm R}_{\earth}$). As in the schematic demonstration in \citet{OwenWu2017}, the initial atmosphere mass fraction $f$ is distributed logarithmically flat between $10^{-5}$ and 1. In Section \ref{sec:obs} we employ a more physically-motivated distribution.

Fig. \ref{fig:simple} clearly illustrates that envelopes heavier than several percent survive (the right peak), while lighter envelopes are blown off completely (the left peak), forming a double-peaked distribution with a valley between $1.5-2.0 \,{\rm R}_{\earth}$. Interestingly, a logarithmically flat initial distribution of $f$ develops a valley in the radius distribution even when the mass loss is turned off (i.e., no photoevaporation and no core-powered loss), and planets are only allowed to cool according to equation \eqref{eq:cooling_eq}. This is because light envelopes are optically thinner and therefore they cool and shrink rapidly, according to equation \eqref{eq:lum}, whereas heavy envelopes cool slowly, roughly retaining their initial radius. By eroding light atmospheres, mass loss due to core cooling amplifies this effect and deepens the valley.  

\subsection{Comparison with observations}\label{sec:obs}

In order to check whether our evolutionary tracks can reproduce more specific features of the observed distribution, we simulate more realistic planet populations.

\begin{figure}
	\includegraphics[width=\columnwidth]{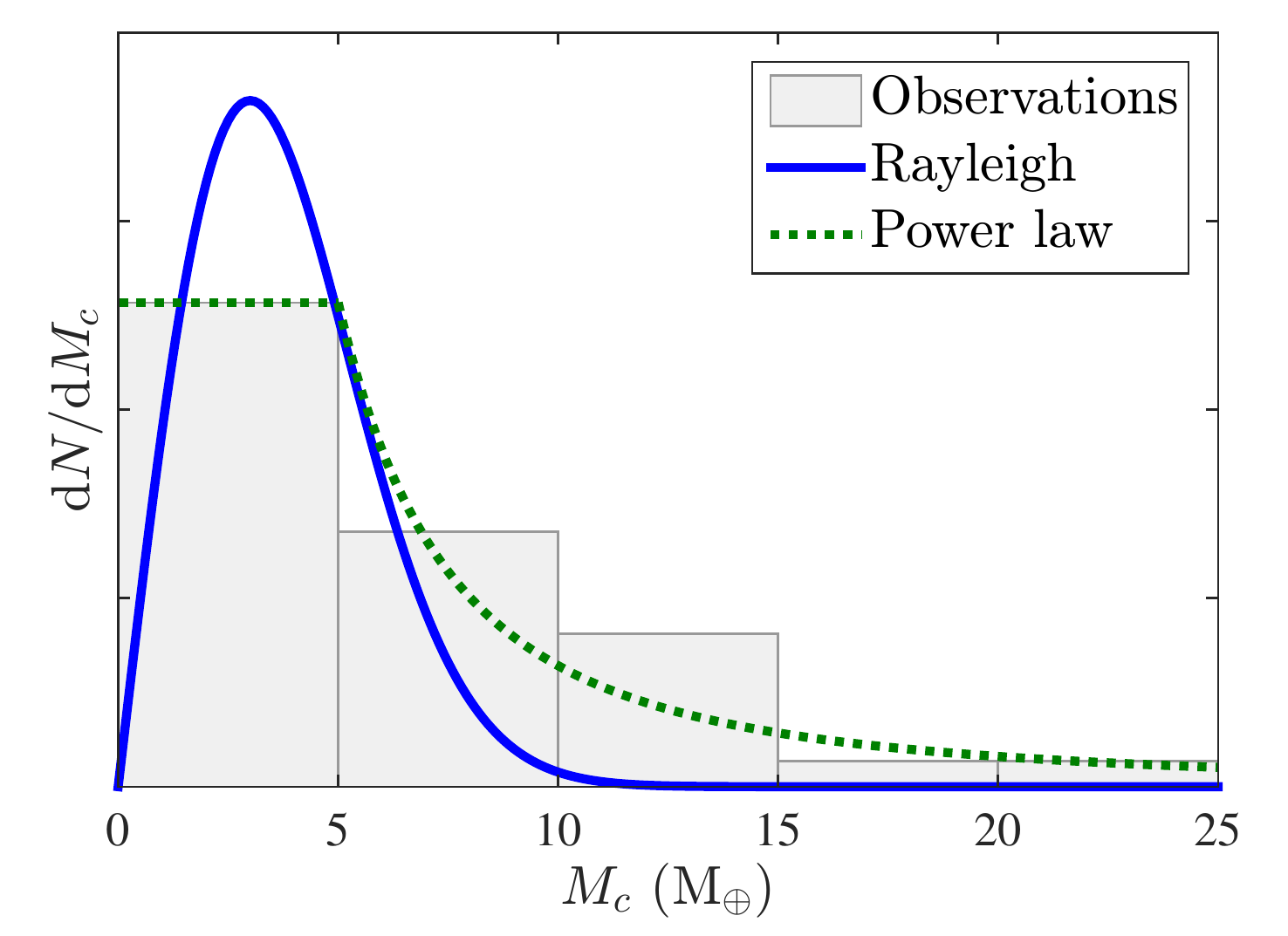}
	\caption{Core-mass distributions that are used in this work. The solid blue line is according to equation \eqref{eq:mass_dist} and it generates the solid blue radius distribution in Fig. \ref{fig:panels} (middle panel). It is also used by \citet{OwenWu2017}. The dotted green line is according to equation \eqref{eq:mass_power_law} and it generates the dotted green radius distribution in Fig. \ref{fig:panels} (bottom panel). The observed distribution (grey histogram) is from \citet{Marcy2014_apjs}.}
	\label{fig:marcy}
\end{figure}

First, we take into account that planets are found at different orbital periods $P\propto T_{\rm eq}^{-3}$, and therefore have a variety of equilibrium temperatures. According to the discussion below equation \eqref{eq:rho_rb}, we expect spontaneous mass loss to be less efficient for colder planets. For comparison with \cite{OwenWu2017}, we roughly follow their orbital period distribution and distribute our temperatures according to
\begin{equation}\label{eq:tmp_dist}
\frac{{\rm d}N}{{\rm d\log}(T_{\rm eq})}\propto\begin{cases}
\textrm{constant} & 500\textrm{ K}<T_{\rm eq}<1000\textrm{ K}\\
T_{\rm eq}^{-6} & 1000\textrm{ K}<T_{\rm eq}<2000\textrm{ K}\,.
\end{cases}
\end{equation}
Next, we distribute our core masses according to a broken power-law distribution
\begin{equation}\label{eq:mass_power_law}
\frac{{\rm d}N}{{\rm d}M_c}\propto\begin{cases}
\textrm{constant} & M_c<5\,{\rm M}_{\earth}\\
M_c^{-2} & M_c>5\,{\rm M}_{\earth}\,.
\end{cases}
\end{equation}
\citet{OwenWu2017}, on the other hand, use a Rayleigh distribution 
\begin{equation}\label{eq:mass_dist}
\frac{{\rm d}N}{{\rm d}M_c}\propto M_c\exp\left(-\frac{M_c^2}{2\sigma_M^2}\right)
\end{equation}
with $\sigma_M=3{\rm M}_{\earth}$. Fig. \ref{fig:marcy} shows that our distribution is more consistent with the observed high-mass tail from radial velocity measurements \citep{Marcy2014_apjs}. 
Unlike \cite{OwenWu2017}, who use a logarithmically flat distribution for the initial atmosphere mass fraction, we correlate $f\propto M_c^{1/2}$, according to theoretical planet formation models \citepalias{Ginzburg2016}\footnote{The relevant initial condition for the thin regime is given by equation (24) of \citetalias{Ginzburg2016}, which accounts for both atmosphere accretion and the mass loss during the thick-atmosphere regime. Adopting different opacities and values of $\gamma$ \citep[as in][who, however, do not consider mass loss]{LeeChiang2015} would have resulted in a somewhat different power-law correlation. Nevertheless, our results are not sensitive to the exact choice of this power.}.

\begin{figure}
	\includegraphics[width=\columnwidth]{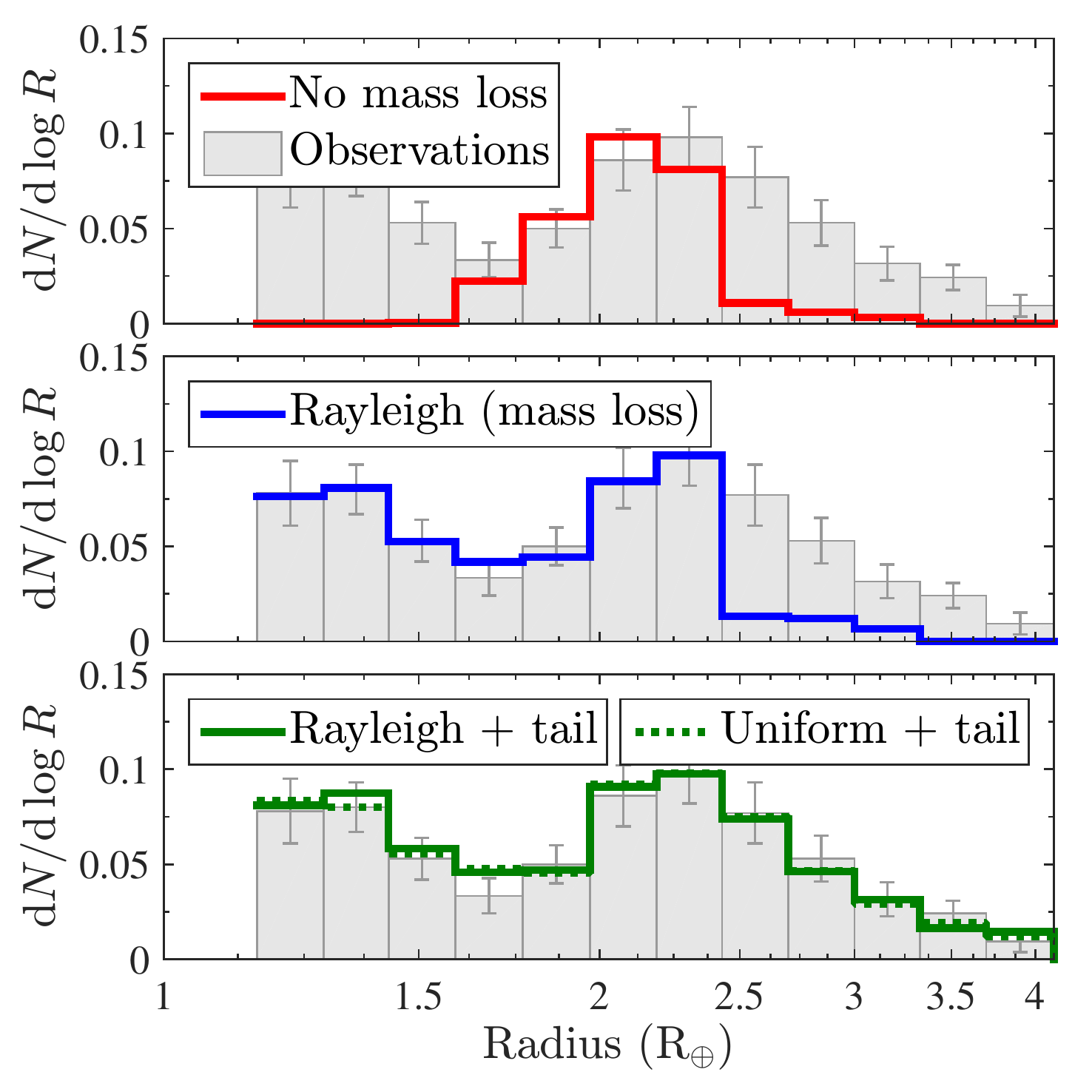}
	\caption{Distribution of planet radii after 3 gigayears of evolution (starting from $R=2R_c$). The equilibrium temperature is distributed according to equation \eqref{eq:tmp_dist} and the initial $f\equiv M_{\rm atm}/M_c$ is given by $f=0.05(M_c/{\rm M}_{\earth})^{1/2}$. The observed distribution (grey histogram) is from \citet{Fulton2017}. {\it Top:} the core mass is distributed according to the Rayleigh distribution of equation \eqref{eq:mass_dist} and the mass loss is turned off. {\it Middle:} same as the top panel, but with the mass loss turned on. {\it Bottom:} same as the middle panel, but with the Rayleigh distribution replaced by a ${\rm d}N/{\rm d}M_c\propto M_c^{-2}$ power law for $M_c>5{\rm M}_{\earth}$ (solid) or with the whole distribution given by equation \eqref{eq:mass_power_law} (dotted).}
	\label{fig:panels}
\end{figure}

In Fig. \ref{fig:panels} we present the results of over 6,500 evolutionary tracks and compare them to the observed distribution from \citet{Fulton2017}. Our nominal results ({\it bottom panel, dotted green}) are obtained using the power-law tail of equation \eqref{eq:mass_power_law}. The other panels use the Rayleigh distribution for a pedagogical demonstration. Without mass loss ({\it top panel, red}) only the right peak is reproduced. This peak corresponds to $\approx 3{\rm M}_{\earth}$ cores that retain their atmospheres. When mass loss is taken into account ({\it middle panel, blue}), the light and hot cores lose their atmospheres completely, producing the left peak. In this way, a single population evolves into a double-peaked distribution that closely resembles the observed valley. However, the Rayleigh model (middle panel, blue) suffers from a clear deficit of Neptune-size ($2.5-4\,{\rm R}_{\earth}$) planets. The reason for this is that equation \eqref{eq:mass_dist} underestimates the abundance of massive cores, as evident in Fig. \ref{fig:marcy} \citep[see also][]{OwenWu2017}. When we replace ({\it bottom panel, solid green}) the exponential tail with a power law for $M_c>5{\rm M}_{\earth}$ (continuously concatenated), the observed distribution is reproduced. In fact, since \citet{Fulton2017} include only planets larger than $1.16\,{\rm R}_{\earth}$, and since the Rayleigh distribution is approximately constant in the low-mass range ($1.8-5\,{\rm M}_{\earth}$), we can safely substitute it for simplicity with a uniform distribution, reverting to our nominal equation \eqref{eq:mass_power_law}. 

\begin{figure}
	\includegraphics[width=\columnwidth]{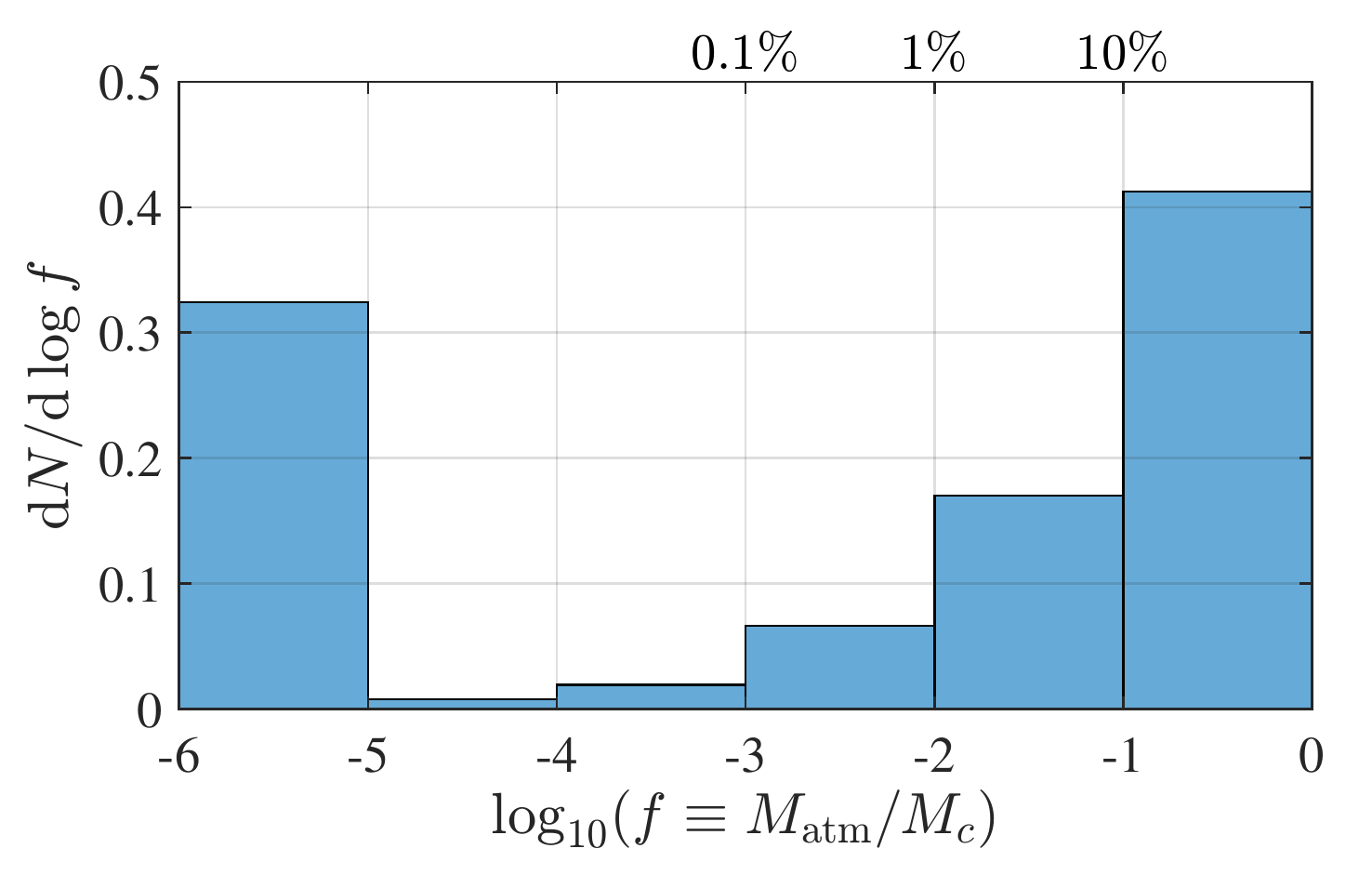}
	\caption{Distribution of the final atmosphere mass fraction $f$ in our nominal model (dotted green line, bottom panel of Fig. \ref{fig:panels}). Atmospheres with $f<10^{-6}$ are considered to be lost completely and are binned in the leftmost bin.}
	\label{fig:f_dist}
\end{figure}

Fig. \ref{fig:f_dist} displays the final atmosphere mass fraction distribution, emphasizing the bimodal nature of the core-powered mass loss.

\subsection{Dependence on orbital period}

\begin{figure}
	\includegraphics[width=\columnwidth]{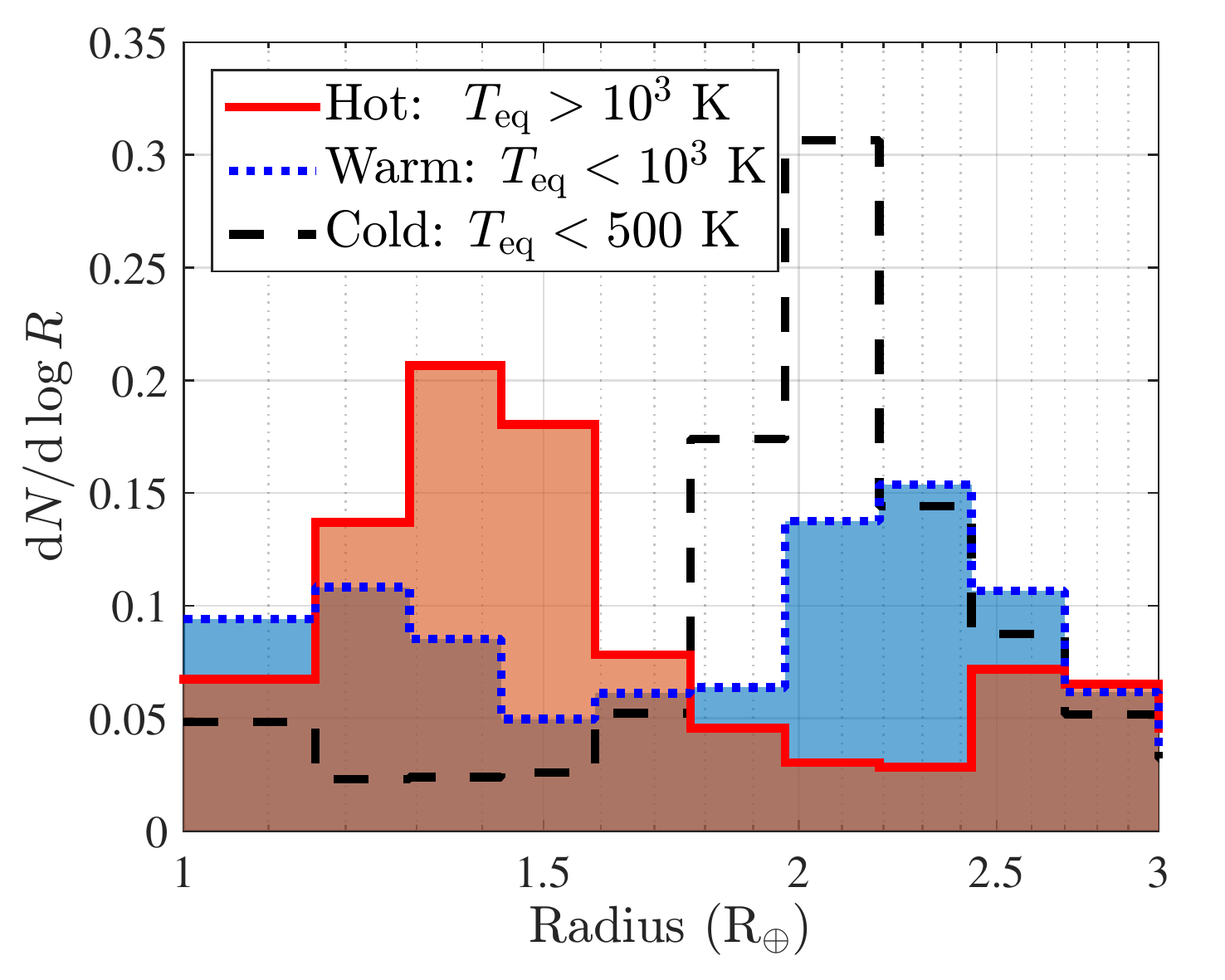}
	\caption{Distribution of the final planet radii in our nominal model (dotted green line, bottom panel of Fig. \ref{fig:panels}), divided into hot (solid red) and warm (dotted blue) populations. The distribution of each group is normalized separately. We also show the distribution of colder planets 300--500 K (unfilled dashed black) which are excluded from \citet{Fulton2017} and Fig. \ref{fig:panels}.}
	\label{fig:hot_cold}
\end{figure}

Since the Bondi condition limits the gas escape (see Section \ref{sec:light}), we expect core-powered mass loss to be more prominent at high equilibrium temperatures, i.e. short orbital periods or hot stars \citepalias[see also][]{Ginzburg2016}. In Fig. \ref{fig:hot_cold} we show how the shape of the valley in the radius distribution changes with $T_{\rm eq}$. Specifically, at temperatures $T_{\rm eq}>10^3\textrm{ K}$ (equivalently, $P\lesssim 10\textrm{ days}$), the distribution is dominated by the left peak, which contains the stripped cores. For $T_{\rm eq}<10^3\textrm{ K}$ ($P\gtrsim 10\textrm{ days}$), the right peak, which is composed of surviving atmospheres, is more prominent as expected. Note that according to equation \eqref{eq:tmp_dist} warm planets are more abundant than hot ones, so the overall distribution (bottom panel of Fig. \ref{fig:panels}) has a somewhat higher right peak.

The observed scatter of radii and orbital periods or incident flux \citep{Lundkvist2016,Fulton2017} shows a similar trend to Fig. \ref{fig:hot_cold} \citep[so does the photoevaporation model; see][]{OwenWu2017}.

\section{Summary and Discussion}\label{sec:summary}

Recent studies identify a valley in the radius distribution of small (a few ${\rm R}_{\earth}$) planets with short (< 100 days) orbital periods \citep{Fulton2017}. Atmosphere erosion by high-energy stellar photons (photoevaporation) has been invoked to explain this valley \citep{OwenWu2017}. 

In a previous paper \citepalias{Ginzburg2016} we explained that spontaneous atmosphere mass loss, powered by the cooling luminosity of the rocky core \citep[see also][]{IkomaHori2012}, can also naturally lead to a bimodal population. Explicitly, if the heat capacity of a gas envelope is larger than that of the underlying rocky core, then it survives roughly intact, significantly increasing the planet's radius. If, on the other hand, the core dominates the heat capacity, then its cooling luminosity may blow off the light atmosphere entirely in a runaway process, reducing the planet's size to that of its bare core.

In this paper we applied our theory to realistic planet populations by evolving them according to a simple analytical prescription that accounts for the simultaneous cooling and spontaneous mass loss (but not stellar photoevaporation) of their gas envelopes. We demonstrated that these populations are naturally transformed by core-powered mass loss into a double-peaked size distribution that closely resembles the observations, including their dependence on the orbital period.

To summarize, core-powered mass loss alone easily reproduces the main features of the observed radius distribution. Since core cooling is a simple process that is regulated by the planet itself and depends on the stellar bolometric luminosity, regardless of the less understood and more variable high-energy tail, it poses an appealing alternative to the standard photoevaporation explanation \citep[][]{MurrayClay2009,OwenJackson2012,OwenAlvarez2016}.

\subsection{Relation to Owen \& Wu (2017)}\label{sec:relation}

This work was motivated by the recent \citet{OwenWu2017} paper. Essentially, both studies evolve fairly similar initial populations of planets and compare the resulting radius distribution to \citet{Fulton2017}.

While the cooling of the planets is treated similarly in both studies \citep[the heat capacity of the core is considered in][]{OwenWu2017}, we list below the key differences between the two papers:
\begin{enumerate}
\item
The removal of the gas envelopes is modelled according to two fundamentally different processes. \citet{OwenWu2017} consider only the high-energy stellar flux as an energy source that unbinds the atmosphere, while we consider only the cooling luminosity of the planet itself \citep[see also][]{OwenWu2016}, and especially of its core. \citet{OwenWu2017} suggest that this spontaneous mass loss can be ignored because it only enhances the mass loss of envelopes that are on their way to be completely stripped anyway by photoevaporation. We, on the other hand, focus on the spontaneous mass loss and demonstrate that it can explain the observed distribution on its own, regardless of the high-energy flux. We suggest, with a similar logic to \citet{OwenWu2017}, that including photoevaporation would strip only light and hot atmospheres, that are on their way to be completely removed by core cooling anyway.
\item
The high-mass tail of our core-mass distribution is consistent with radial velocity measurements (Fig. \ref{fig:marcy}).
\item 
Our initial $M_{\rm atm}$ distribution is consistent with core-accretion models, in contrast to the logarithmically flat distribution of the initial $f\equiv M_{\rm atm}/M_c$ in \cite{OwenWu2017}. Actually, our Fig. \ref{fig:simple} suggests that the valley in the schematic demonstration of \citet{OwenWu2017} is partially a result of their assumed initial conditions of $f$, rather than photoevaporation (the situation is different for their nominal model, in which $f$ is also distributed logarithmically flat, but from a much narrower range).
\end{enumerate} 

\subsection{Outlook}

It is important to distinguish between the different mass loss mechanisms (core cooling versus photoevaporation). However, since both processes seem to behave similarly (including their scaling with the equilibrium temperature, which differs only for high temperatures and masses; see Fig. 3 of \citetalias{Ginzburg2016}), distinguishing between them is difficult. A promising avenue will be to compare the distributions of planets around different stellar types \citep[in this context, see the recent paper by][]{Hirano2017}. This will exploit the main difference between the two models: photoevaporation is powered by the high-energy tail of the stellar irradiation, whereas core cooling correlates to the total bolometric flux (through $T_{\rm eq}$). Another aspect of this difference is the large scatter in the high-energy flux of stars with the same mass \citep{Tu2015}, leading to a less distinct photoevaporation ``desert'' of short-period low-density planets, compared to the core cooling scenario. Future observations could also study the dissimilar early histories of photoevaporation and spontaneous mass loss, which act on different timescales. 

Though our model can be used to infer the initial atmosphere masses from the observed radius distribution and thereby constrain planet formation theories, there is some inherent degeneracy in the initial conditions due to the runaway nature of the mass loss.

\section*{Acknowledgements}

This research was partially supported by ISF (Israel Science Foundation) and iCore (Israeli Centers of Research Excellence) grants. HES gratefully acknowledges support from NASA grant NNX15AK23G.
We thank Yoram Lithwick for discussions during the 2015/6 Jerusalem Winter School. We also thank Fei Dai, Christoph Mordasini, James Owen, Allona Vazan and Josh Winn for discussions and comments on the paper's draft. Finally, we thank Eugene Chiang for a careful review that significantly improved the paper.  




\bibliographystyle{mnras}
\bibliography{core} 




\bsp	
\label{lastpage}
\end{document}